\documentclass[pre,preprint,showkeys,amssymb,amsmath,aps]{revtex4}
\usepackage{amsmath}
\usepackage{amsfonts}
\usepackage[dvips]{graphicx}
\usepackage{epsfig} 
\usepackage[]{subfigure}
\usepackage[dvips]{color}
\usepackage{color}

\usepackage{latexsym} 
\usepackage{amsmath,amscd}
\usepackage{times}

\begin{document}

\title{\lowercase{flat}IGW - an inverse algorithm to compute the Density of States of lattice Self Avoiding Walks}
\author{M. Ponmurugan${}^{+}$ and V. Sridhar${}^{\star}$}
\affiliation{${}^{+}$Institute of Mathematical Sciences, C.I.T. Campus, Taramani Chennai 600 113, India \\ 
${}^{\star}$Materials Science Division, Indira Gandhi Centre for Atomic Research,
Kalpakkam 603102, Tamilnadu, India}

\author{S. L. Narasimhan${}^{\dagger \#}$ and K. P. N. Murthy${}^{\ddagger}$} 
\affiliation{${}^{\dagger}$Solid State Physics Division, Bhabha Atomic Research Centre, 
Mumbai 400085, Maharashtra, India \\
${}^{\ddagger}$School of Physics, University of Hyderabad, Central University P.O.,
Gachibowli, Hyderabad 500 046, Andhra Pradesh, India }

\date{\today}

\begin{abstract}

We show that the Density of States (DoS) for lattice Self Avoiding Walks can be estimated by using an inverse algorithm, called flatIGW, whose step-growth rules are dynamically adjusted by requiring the energy histogram to be locally flat. Here, the (attractive) energy associated with a configuration is taken to be proportional to the number of non-bonded nearest neighbor pairs (contacts). The energy histogram is able to explicitly direct the growth of a walk because the step-growth rule of the Interacting Growth Walk \cite{IGW} samples the available nearest neighbor sites according to the number of contacts they would make. We have obtained the complex Fisher zeros corresponding to the DoS, estimated for square lattice walks of various lengths, and located the $\theta$ temperature by extrapolating the finite size values of the real zeros to their asymptotic value, $\sim 1.49$ (reasonably close to the known value, $\sim 1.50$ \cite{barkema}). 

\end{abstract}


\maketitle

\section{Introduction}

One of the challenging computational problems in the study of Self-Avoiding Walks (SAW) \cite{vander} - a statistical mechanical model for describing the excluded volume interaction of a linear polymer chain - is to estimate the configurational Density of States (DoS) as accurately as possible so that their equilibrium thermodynamic properties can be worked out. Equivalently, the problem is to find an efficient method for generating all possible configurations of a SAW. Since the total number of SAW configurations increases exponentially with the length, $N$, of the walk, there is a need to devise efficient sampling techniques for generating as wide a spectrum of configurations as possible. 

The standard unbiased methods of generating SAW configurations \cite{sokal}, on a regular lattice, not only suffer from sample-attrition problem but also fail to generate low-entropy configurations in sufficient numbers. In other words, an unbiased method generates only typical SAW configurations with equal {\it a priori} probability; low-entropy (or compact) configurations are rarely generated because of their very low probability of occurrence in the microcanonical ensemble of all possible configurations. So, Kinetic Growth Walk (KGW) \cite{KGW} was proposed as a method for introducing local bias in the generation process so that the typical SAW configurations generated are more compact than the ones generated by the unbiased method; the sample-attrition problem also is less severe. The probability of generating a given KGW configuration depends on the way it is grown step-by-step. Since the cumulative bias introduced for generating a successful KGW configuration is known, it can be removed by suitably weighting the configuration.

Recently, Prellberg and Krawczyk \cite{flatperm} have shown that the KGW algorithm is actually an approximate counting algorithm in which the weights associated with the walk configurations are the estimates of the number of configurations. During the generation process, decision to stop or continue the walk (a certain number of times) is based on how the weight, $a_N$, associated with it compares with the average of weights, ${\bar a}_N$, associated with all the successful, as well as failed, configurations thus far generated. After a sufficiently large number of trials, ${\bar a}_N$ gives a fairly accurate estimate of the microcanonical partition function for SAW. This algorithm, known as 'flatPERM', can be used for estimating the DoS if the configurations are binned with respect to the number of {\it contacts} ({\it i.e}., non-bonded nearest neighbor pairs) they make. A reasonably flat {\it contacts}-histogram, over the full range of contacts possible for a given value of $N$, provides an independent numerical evidence that rare configurations are not missed.

An interesting inverse problem is to find out whether the requirement of a flat energy-histogram (or equivalently, {\it contacts}-histogram) could lead to an accurate estimate of the DoS. We show that such an inverse method can be devised and used for obtaining the DoS, taking the Interacting Growth Walk (IGW) algorithm as an example. IGW \cite{IGW} is a finite temperature ($\beta _G^{-1}$) generalization of the KGW that grows by Boltzmann-sampling the locally available sites according to the number of contacts (or, 'energy') they would make. We show that the inverse method dictates adjusting the parameter, $\beta _G^{-1}$, which is equivalent to adjusting the weight of the growing configuration. We compute the Fisher zeros \cite{Yang, finsy,pzprivman,Korean} corresponding to the DoS, estimated for SAW on a square lattice, and locate the $\theta$-point. Agreement with the expected value of the transition temperature justifies the idea of an inverse algorithm which is similar in spirit to the Reverse Monte Carlo (RMC) algorithm of McGreevy and Pusztai \cite{McGreevy}. 

The paper is organized as follows. In the next section, we briefly review the IGW model for generating lattice SAWs whose compactness can be tuned by a parameter. We point out that averaging the bias-correcting weights over all attempts to generate walks of given length gives an estimate of the microcanonical partition function. In the first part of section III, using the examples of KGW and IGW on a honeycomb lattice, we show that the growth rule of flatPERM consists of two independent parts - (i) a {\it local} KGW rule and (ii) a {\it global} weight-based rule for continuing a walk; in particular, we show that the weight-based rule cannot in general be reinterpreted in terms of a 'bath' temperature. 
In the second part of this section, we define the inverse algorithm, flatIGW, whose step-growth rules are dynamically adjusted by the requirement of local flatness for the energy histogram. In section IV, we present the Monte Carlo estimates of the DoS obtained for SAWs on a square as well as triangular lattice. We also present an extrapolation analysis of the Fisher zeros obtained for square lattice SAWs. In the last section, we summarize our results. 

\section{Interacting Growth Walk (IGW)}

Starting from an arbitrary site of a regular lattice of coordination number, $z$, the first step is taken in one of the available $z$ directions; choice for the subsequent steps is made from among the $z-1$ nearest neighbor sites in the forward direction. In the unbiased sampling method, an attempt to generate a walk is discarded if the site chosen has been visited already. Even though this method ensures that the successful configurations are all generated with equal probability, $P_N^{SAW} = z(z-1)^N$, the rejection rate increases exponentially with the length of the walk because of which it fails to generate sufficient number of successful walks needed for good statistical analysis.

An immediate solution to this problem is to choose at random one of the available, or unvisited, nearest neighbors for the next step to be taken. The number of available sites for the $N^{th}$ step, $a_N({\cal C}_{N-1})$ ($\in [0,z-1]$), depends on the configuration ${\cal C}_{N-1}$ already generated. It is clear that this method, known as the KGW, introduces bias at every step of the walk. However, sample-attrition is less severe in this case and is only due to the geometrical 'trapping' ($a_N({\cal C}_{N-1}) = 0$) of the walk. The cumulative bias introduced during the generation process can be removed by associating a weight, $A_N^{KGW}({\cal C}_N) = z\times \prod _{K=2}^N a_K({\cal C}_{K-1})$, with a configuration. These weights, referred to as the Rosenbluth-Rosenbluth (RR) weights \cite{KGW}, may be used for estimating the mean squared end-to-end distance,
\begin{equation}
\langle r_N^2 \rangle = \frac{\sum _{{\cal C}_N} A_N^{KGW}({\cal C}_N)r_N^2({\cal C}_N)}{\sum _{{\cal C}_N} A_N^{KGW}({\cal C}_N)}
\label{eq:msd}
\end{equation}
As $N$ becomes larger, the distribution of these RR-weights becomes flatter and so, the weighted averages fail to provide  bias-corrected estimates in the asymptotic limit of very long walks. 

Averaging the RR-weights over all the configurations, trapped or successful, we obtain an estimate of the total number of $N$-step SAWs \cite{flatperm}:
\begin{equation}
{\tilde Z}_N^{SAW} = \frac{1}{{\cal S}}\sum _{{\cal C}_N}A_N^{KGW}({\cal C}_N)
\label{eq:Zsaw1} 
\end{equation}
where ${\cal S}$ is the total number of attempts made by the KGW algorithm to generate $N$-step SAWs; since the weights associated with those configurations that have been trapped at various stages are zero, the summation is only over all the successful configurations. ${\tilde Z}_N^{SAW}$ will give the total number (equivalently, the microcanonical partition function) of $N$-step SAWs, $Z_N^{SAW}$, in the limit ${\cal S}\to \infty$.

Since any given KGW configuration, ${\cal C}_N$, will have a certain number of non-bonded nearest neighbor pairs ({\it contacts}), we can equivalently define,
\begin{equation}
{\tilde Z}_N^{SAW} = \sum _{m=1}^{m_x} a_{N,m}; \quad 
             a_{N,m} \equiv \frac{1}{{\cal S}}\sum _{{\cal C}_N} A_N^{KGW}({\cal C}_N,m) 
\label{eq:Zsaw2} 
\end{equation}
where $A_N^{KGW}({\cal C}_N,m)$ is the RR-weight associated with a configuration that has $m$ contacts. In the limit ${\cal S}\to \infty$, therefore, $a_{N,m}$ is the total number of $N$-step SAWs with $m$ contacts, which is the DoS. It must be noted that the KGW algorithm is blind to the contacts being made during growth - {\it i.e.,} identifying and counting the contacts are not a part of the basic growth rule. 

On the other hand, IGW algorithm \cite{IGW} samples the available nearest neighbors on the basis of the number of contacts they would make. The probability that an available nearest neighbor site, say $r_N(l)$, for the $N^{th}$ step will be chosen is given by
\begin{equation}                              
p_N^{IGW}({\cal C}_{N-1},\beta _G) \equiv \frac{e^{-\epsilon \beta _G m(r_N(l))}}{a_N({\cal C}_{N-1},\beta _G)}; \quad
a_N({\cal C}_{N-1},\beta _G) \equiv {\sum _{l=1}^{a_N({\cal C}_{N-1})}e^{-\epsilon \beta _G \ m(r_N(l))}}
\label{eq:pIGW}
\end{equation}
where $m(r_N(l))$ is the number of contacts the site $r_N(l)$ would make, and $\epsilon$ is the 'energy' cost of a contact. The parameter $\beta _G$ is an inverse 'temperature' that helps sample the local environment. For $\beta _G = 0$, this is just the KGW growth rule; for $\beta _G \to \infty$, only the site making a maximum number of contacts will be chosen. Interestingly, there is no sample-attrition in the limit $\beta _G \to \infty$ because a {\it contact}-making site cannot be avoided during growth, and the walks thus generated are maximally (not globally) compact. Even though $\beta _G^{-1}$ cannot be identified with a canonical 'bath' temperature \cite{igwhoney}, it has been shown \cite{IGW} that a coil-to-globule transition can be identified for some lattice-dependent value of $\beta _G^{-1}$.

As in the case of KGW, a weight $A_N^{IGW}({\cal C}_N,\beta _G)$ can be associated with a given IGW configuration,
\begin{equation}
A_N^{IGW}({\cal C}_N,\beta _G) = z \times \prod _{K=2}^N a_N({\cal C}_{N-1},\beta _G)
\label{eq:wtIGW}
\end{equation}  
so that an estimate of the microcanonical partition function for SAW is given by the average,
\begin{equation}
{\tilde Z}_N^{SAW} = \frac{1}{{\cal S}(\beta _G)}\sum _{{\cal C}_N}A_N^{IGW}({\cal C}_N,\beta _G)
                     e^{\epsilon \beta _G \ m({\cal C}_N)}
\label{eq:Zsaw3} 
\end{equation}
where ${\cal S}(\beta _G)$ is the total number of attempts made by the IGW algorithm to generate $N$-step SAWs, and $m({\cal C}_N)$ is the number of contacts in ${\cal C}_N$. In the limit ${\cal S}(\beta _G) \to \infty$, the microcanonical partition function obtained this way is independent of the parameter $\beta _G$. The DoS is then given by 
\begin{equation}
a_{N,m} = \lim _{{\cal S}(\beta _G)\to \infty }\frac{1}{{\cal S}(\beta _G)}\sum _{{\cal C}_N}A_N^{IGW}({\cal C}_N,\beta _G,m)
                                               e^{\epsilon \beta _G \ m({\cal C}_N)}
\label{eq:dos}
\end{equation}
where $A_N^{IGW}({\cal C}_N,\beta _G,m)$ is the weight associated with an IGW configuration ${\cal C}_N$ with $m$ contacts. For $\beta _G = 0$, it is just the RR-weight. On the other hand, averaging only the weights, $A_N^{IGW}({\cal C}_N,\beta _G)$, 
\begin{equation}
{\tilde Z}_N^{SAW} = \frac{1}{{\cal S}(\beta _G)}\sum _{{\cal C}_N}A_N^{IGW}({\cal C}_N,\beta _G)
\label{eq:Zsaw4} 
\end{equation}
leads to the canonical partition function at the 'bath' temperature $\beta _G^{-1}$ in the limit ${\cal S}(\beta _G) \to \infty$.

\section{\lowercase{flat}IGW - an inverse algorithm}

As mentioned earlier, a typical KGW configuration is more compact than a typical SAW configuration because it is generated by sampling only sites available for growth with the same probability; this introduces a local bias not because it distinguishes between the available sites but because the number of available sites is random and configuration-dependent. flatPERM uses the RR-weights to decide whether the walk should be continued and with what adjustment to its RR-weight. So, the growth rule consists of two independent parts - (i) {\it local} (equivalently, the KGW) sampling of the available nearest neighbors with equal probability, and (ii) {\it global} check on how the RR-weight of the currently growing configuration compares with a cumulative average. The {\it local} (KGW) growth rule is independent of the {\it global} weights-based rule and, is never modified. The objective of this section is to use the solvable models of KGW and IGW on a honeycomb lattice in order to demonstrate that this independence is mutual.

\subsection{\lowercase{flat}PERM {\it versus} step-growth rule}

\noindent (1) \underline{KGW:}

Poole {\it et al.,} \cite{poole} have shown that KGW on a honeycomb lattice is equivalent to the Interacting Self-Avoiding Walk (ISAW) at the inverse 'bath' temperature $\beta _{K(hc)} \equiv \ln 2$. This is so because, on a honeycomb lattice, a walk makes a contact if only one nearest neighbor site is available for further growth. Therefore, the probability of generating a walk, ${\cal C}_N$, with $m({\cal C})$ contacts is given by
\begin{equation}
P_N^{KGW}({\cal C}_N) = \frac{1}{3}\times \frac{1}{2^{N-1}}\times 2^m \equiv P_N^{SAW}\times e^{-\epsilon m({\cal C}_N)\ \beta _{K(hc)}}
\label{eq:pKGWhc}
\end{equation} 
where each contact is assigned energy $\epsilon = -1$. It is clear that KGW is equivalent to ISAW at a 'bath' temperature $\beta _{K(hc)}^{-1} =  (\ln 2)^{-1}$. Since the 'bath' temperature is higher than the $\theta$ temperature on a honeycomb lattice ($\sim 1$), KGW is in the same universality class as SAW and therefore, {\it unweighted} averaging of the end-to-end distances of KGW also yields the SAW value for the size-exponent $\nu = 3/4$. However, configuration-dependence of $P_N^{KGW}({\cal C}_N)$ necessitates weighting the configurations appropriately so that they represent a {\it bona fide} statistical ensemble.  

The (microcanonical) weight associated with a KGW configuration, ${\cal C}_N$, is then given by
\begin{equation}
A_N^{K(hc)}({\cal C}_N) = (P_N^{KGW}({\cal C}_N))^{-1} 
                        = \frac{e^{\epsilon m\ \beta _{K(hc)}}}{P_N^{SAW}}
\label{eq:wKGWhc}
\end{equation}
This is just the RR-weight for a KGW configuration because $\beta _{K(hc)} = \ln 2$ corresponds to the athermal case on a honeycomb lattice. The flatPERM parameter, $r_N$, is then defined as the ratio
\begin{equation}
r_N = \frac{A_N^{K(hc)}({\cal C}_N)}{{\bar A}_N^{K(hc)}}; \quad 
{\bar A}_N^{K(hc)} \equiv \frac{1}{{\cal S}}\sum _{{\cal C'}_N}A_N^{K(hc)}({\cal C'}_N) 
\label{eq:rKGWhc1}
\end{equation}
where ${\cal S}$ is the total number of attempts made (successful or failed). Since $P_N^{SAW}$ is a constant, we have the ratio,
\begin{equation}
r_N = \frac{e^{\epsilon m({\cal C}_N)\ \beta _{K(hc)}}}{{\bar a}_N^{K(hc)}}; \quad
{\bar a}_N^{K(hc)} \equiv \frac{1}{{\cal S}}\sum _{{\cal C'}_N}e^{\epsilon m({\cal C'}_N)\ \beta _{K(hc)}} 
\label{eq:rKGWhc2}
\end{equation}
If $r_N < 1$, the walk will be continued with probability $r_N$ after resetting its weight to be equal to ${\bar a}_N^{K(hc)}$. This situation is realized when the current configuration is more compact (with $\epsilon = -1$) than an average or typical configuration. On the other hand, if $r_N > 1$, it is copied $c =$ min$([r_N],z_a)$ times, and each copy continued after reducing its weight by $c$; here, $[r_N]$ is the largest integer less than $r_N$ and $z_a (= 1,2)$ is the number of available sites. In this case, the walk is less compact (with $\epsilon = -1$) than a typical configuration and therefore can be reused;

Resetting the weight, $a_N^{K(hc)} \to a_N^{'K(hc)} = {\bar a}_N^{K(hc)}$ or $a_N^{K(hc)}/c$, is formally equivalent to  resetting the inverse temperature $\beta _{K(hc)}$ to a value given by
\begin{equation}
\beta _{K(hc)} \to \beta _{'K(hc)} = \frac{1}{\epsilon \ m({\cal C}_N)}\ln a_N^{'K(hc)}
\label{eq:dbeta1}
\end{equation}
This only provides an alternative interpretation of the {\it global} weight-based rule for a chain-growth process because of the known fact that the  'bath' temperature decides the degree of compactness of a typical configuration. However, this cannot be used as a parameter in the {\it local} growth rule because there is nothing to distinguish one available nearest neighbor from another for taking a given step. 

On a lattice of coordination number $z>3$, even a configuration-independent bath temperature for KGW cannot be defined \cite{igwhoney}. For example , on a square lattice ($z=4$), a walk makes a single (or double) contact if the site chosen has one (or two) occupied non-bonded nearest neighbor. Note that the walk is geometrically trapped if it chooses a site having three occupied non-bonded nearest neighbors. It is clear that the number of single ($m_1)$  and double $(m_2)$ contacts made is configuration-dependent. Averaging over all the configurations having the same total number ($m$) of contacts, but with different sets of single and double contacts, $\{ (m_1,m_2) \mid m = m_1+m_2\}$, we can define an $m$-dependent inverse temperature,
\begin{equation}
\beta (m) \equiv \frac{{\bar m}_1 \ln (3/2) + 2{\bar m}_2 \ln 3}{m}; \quad \beta (0) = 0
\label{eq:betam}
\end{equation}
where ${\bar m}_{1,2}$ denotes configurational averages. Since these $\beta$-values are sharply distributed about the value $\beta _{sq} \sim 0.45$ (width $\sim 0.02$) in the asymptotic limit ($N \to \infty$) \cite{igwhoney}, KGW may be said to be equivalent to ISAW at an {\it effective} bath temperature, $\beta _{sq} \sim 0.45$; but the converse is not true. If, however, the unweighted averaging of the end-to-end distances for KGW also gives the SAW value for $\nu (= 3/4)$ \cite{KGW}, it is because $\beta _{sq}^{-1}$ is higher than the $\theta$ temperature ($\sim 1.15$). Hence, even reinterpreting the {\it global} weight-based rule in terms of a 'bath' temperature is not straightforward for walks on a lattice of coordination number $z > 3$. 

\bigskip

\noindent (2) \underline{IGW:}

Since an IGW samples the available nearest neighbors on the basis of the number of contacts they would make, the individual steps of the walk on a honeycomb lattice may be identified either as {\it two-indexed} steps or as {\it forced} steps \cite{igwhoney}. A two-indexed step, labeled $[m_1,m_2]$, is a step to an available site making $m_1$ contacts rather than to the other available site making $m_2$ contacts; on a honeycomb lattice,  $m_1, m_2 = 0, 1, 2$. Steps labeled $[2,0], [2,1]$ and $[2,2]$ indicate that the walk is geometrically trapped. Forced steps are those that have no other choice but to take the only site available; they are preceded either by contact-making two-indexed steps or by forced steps.

The probability of generating a walk, ${\cal C}_N$, with $m({\cal C}_N)$ contacts is given by
\begin{eqnarray}
P_N^{IGW}({\cal C}_N) & = & P_N^{SAW}\times e^{-\epsilon m({\cal C}_N)\ \beta _{K(hc)}}
                                 \times e^{-\epsilon n_{[1,0]}\beta _G} \nonumber \\
                      &   & \times \left( \frac{2}{1+e^{-\epsilon \beta _G}} \right)^
                                                           {n_{[0,1]}+n_{[1,2]}+n_{[1,0]}}  
                            \times \left(\frac{2}{1+e^{-2\epsilon \beta _G}}\right)^{n_{[0,2]}}                                
\label{eq:pIGWhc1}
\end{eqnarray}
where $\beta _G$ is the inverse growth temperature that helps choosing an available nearest neighbor that makes a certain number of contacts; $n_{[m_1,m_2]}$ is the number of steps that have chosen to make $m_1$ rather than $m_2$ contacts. It is quite possible that different configurations with the same number of contacts $m$ may have different sets of numbers $\{ n_{[m_1,m_2]}\}$, we may use their average values, $\{ {\bar n}_{[m_1,m_2]}\}$ and define an inverse temperature,
\begin{eqnarray}
\beta _{I(hc)}(\beta _G, m) & \equiv & \beta _{K(hc)} + \frac{{\bar n}_{[1,0]}}{m({\cal C}_N)}\beta _G -
\frac{({\bar n}_{[0,1]}+{\bar n}_{[1,2]}+{\bar n}_{[1,0]})}{\epsilon m({\cal C}_N)}
\ln \left( \frac{2}{1+e^{-\epsilon \beta _G}}\right) - \nonumber \\
& &\qquad \qquad \frac{{\bar n}_{[0,2]}}{\epsilon m({\cal C}_N)}
\ln \left( \frac{2}{1+e^{-2\epsilon \beta _G}}\right)
\label{eq:bIGW}
\end{eqnarray}
We may now rewrite Eq.(\ref{eq:pIGWhc1}) in the form,
\begin{equation}
P_N^{IGW}({\cal C}_N) = P_N^{SAW}\times e^{-\epsilon m({\cal C}_N)\ \beta _{I(hc)}(\beta _G,m)}
\label{eq:pIGWhc2}
\end{equation}
For a given value of $N$ and $\beta _G$, the distribution of the inverse temperature $\beta _{I(hc)}(\beta _G,m)$ over a large collection of IGW configurations is sharply peaked around an average value, say ${\bar \beta}_{I(hc)}(\beta _G)$, that may be taken to be the bath temperature. Unlike in the case of KGW, an implicit, configuration-independent, bath temperature does not exist for IGW ($\beta _G > 0$) on a honeycomb lattice. 

In order to remove the bias introduced in the walk-generation process, we associate the weight $A_N^{I(hc)}(\beta _G,{\cal C}_N)$ with configuration ${\cal C}_N$,
\begin{equation}
A_N^{I(hc)}(\beta _G,{\cal C}_N) = \frac{e^{\epsilon m({\cal C}_N)\ \beta _{I(hc)}(\beta _G,\ m({\cal C}_N))}}{P_N^{SAW}}
\label{eq:wIGWhc}
\end{equation}
and compute the flatPERM parameter, $r_N$, defined as
\begin{equation}
r_N = \frac{e^{\epsilon m({\cal C}_N)\ \beta _{I(hc)}(\beta _G,\ m({\cal C}_N))}}{{\bar a}_N^{I(hc)}(\beta _G)}; \quad
{\bar a}_N^{I(hc)}(\beta _G) \equiv \frac{1}{{\cal S}}\sum _{{\cal C'}_N}e^{\epsilon m({\cal C'}_N)\ \beta _{I(hc)}(\beta _G,\ m({\cal C'}_N))} 
\label{eq:rIGWhc2}
\end{equation}
Again, depending whether $r_N <$ or $\geq 1$, the walk will be continued with a probability equal to $r_N$ or unity respectively; the weight $a_N^{I(hc)}(\beta _G)$ will be correspondingly reset to the value $a_N^{'I(hc)}(\beta _G) = {\bar a}_N^{I(hc)}(\beta _G)$ or $a_N^{I(hc)}(\beta _G)/c$ where $c = \mbox{min}(z,[r_N])$. Formally, this is equivalent to resetting $\beta _{I(hc)}(\beta _G,\ m({\cal C}_N))$ to a value,
\begin{equation}
\beta' _{I(hc)}(\beta _G,m({\cal C}_N)) = \frac{1}{\epsilon m({\cal C}_N)}\ln a_N^{'I(hc)}(\beta _G)
\label{eq:dbeta3}
\end{equation}
Since the partitioning of $N$ into the number of steps, $n_{[m_1,m_2]}$, that have chosen to make $m_1$ rather than $m_2$ contacts ($N = n_{[0,1]} + n_{[1,0]} + n_{[1,2]}$ on a honeycomb lattice) is not unique for a configuration ${\cal C}_N$ having a total of $m({\cal C}_N)$ contacts, reinterpreting the {\it global} weight-based rule in terms of a 'bath' temperature is not possible even on a honeycomb lattice.

However, by adjusting the value of $\beta _G$ continuously during the growth process, we may be able to generate the desired SAW ensemble because, the {\it local} IGW rule distinguishes the available nearest neighbor sites on the basis of the number of contacts they would make. What is required is a {\it global} criterion that dictates appropriate adjustments to the value of $\beta _G$ at every step. The fact that a flat energy-histogram is an indicator of an efficient DoS algorithm raises the question whether a requirement of local flatness for an energy-histogram, which is building up during the simulation, can be used as a {\it global} criterion for tuning  $\beta _G$. 

\subsection{\lowercase{flat}IGW}

Let $H_K(m_K)$ denote the number of $K$-step IGWs having $m_K$ contacts ($m_K = 0, 1, 2, \cdots , m_K^{(x)}$ where $m_K^{(x)}$ is the maximum number of contacts possible for a $K$-step walk). A perfectly flat energy histogram corresponds to having the same value $H_K(m_K)$ for all $m_K$. If $H_K(m_K) \neq H_K(m_K-1)$, the number of $K$-step walks having $m_K$ contacts is more, or less, than those having $(m_K-1)$ contacts. So, the difference, $\delta H_K(m_K) \equiv H_K(m_K) - H_K(m_K-1)$, may be taken as a measure of the deviation from local flatness. 

Suppose, a $K$-step walk segment has already made $m_K$ contacts. Since it is an irreversibly growing walk, the number of contacts, $m_K$, cannot decrease during further growth; it can either increase or remain the same. Before taking the $(K+1)^{th}$ step, we should check whether it is desirable, or not, to increase the number of contacts. We assume, for simplicity, that no available site increases $m_K$ by more than one. 

Since $\beta _G$ is the parameter that controls the extent to which a contact-making site is preferred, or avoided, its value for the $(K+1)^{th}$ step may be chosen so that $\mid \delta H_{K+1}(m_K+1)\mid$ does not increase. For example, if $\delta H_{K+1}(m_K) \geq 0$, $\beta _G(K+1)$ must have a negative value so that the {\it local} IGW growth-rule avoids a contact-making site; on the other hand, if $\delta H_{K+1}(m_K) < 0$, $\beta _G(K+1)$ must have a positive value so that a contact-making site is preferred. The actual value of $\beta _G(K+1)$ should depend on $\mid \delta H_{K+1}(m_K+1)\mid$. The simplest choice is to have $\mid \beta _G(K+1)\mid \ \propto \ \mid \delta H_{K+1}(m_K+1)\mid$, with the proportionality factor, say $\xi$, chosen conveniently by trial and error. We have taken $\xi$ to be a uniform random number in the unit interval \cite{pon2} and tried the following rule:
\begin{equation}
\mid \beta _G^{(K+1)} \mid \equiv \mbox{min}\{ \mid \delta H_{K+1}(m_K+1)\xi \mid , \mid \beta _G^{max}\mid \}
\label{eq:delta}
\end{equation}
where $\mid \beta _G^{max}\mid $ is the maximum value conveniently chosen to be within the overflow/underflow limits.
The actual value of $\beta _G^{(K+1)}$ used for the $(K+1)^{th}$ step is given by
\begin{eqnarray} 	  	 
 \beta_G^{(K+1)} = \left\{
                \begin{array}{lll}
                 +\mid \beta _G^{(K+1)} \mid &  if  &  \Delta < 0    \\
                 -\mid \beta _G^{(K+1)} \mid &  if  &  \Delta > 0 
                  \qquad \mbox{where} \quad \Delta \equiv \delta H_{K+1}(m_K+1) \ \xi   \\
                  \quad 0 &  if  &  \Delta = 0.  \\
                \end{array} 
                 \right.
\label{eq:dbeta4}    
\end{eqnarray}
Hence, the step-growth probability defined in Eq.(\ref{eq:pIGW}) for constant $\beta _G$ has to be redefined as
\begin{equation}
p_{K+1}^{IGW} = \frac{e^{\beta _G^{(K+1)}m({\vec r}_{K+1})}}
                        {\sum _{k=1}^{a_{K+1}} e^{\beta _G^{(K+1)}m({\vec r}_{K+1}(k))}}
\label{eq:pflatIGW}
\end{equation}
where ${\vec r}_{K+1}$ is the site taken by the $(K+1)^{th}$ step, $m({\vec r}_{K+1})$ is the number of contacts made by occupying that site (assumed to be either 0 or 1 for simplicity) and $a_{K+1}$ is the number of sites available for the ${K+1}^{th}$ step. The procedure, presented above, can easily be rewritten for the case when there is also an available site that increases $m_K$ by more than one.

Depending on whether a contact is made or not at the $(K+1)^{th}$ step, $H_{K+1}(m_K+1)$ or $H_{K+1}(m_K)$ is incremented by one count. Simultaneously, we have a variable ${\bar g}_{K+1}(m_K+1)$ or ${\bar g}_{K+1}(m_K)$ representing the cumulative weights of all the $(K+1)$-step configurations generated. The weights, essentially defined by Eq.(\ref{eq:pIGW}) and Eq.(\ref{eq:wtIGW}), have to reflect the fact that the growth parameter $\beta _G(K+1)$ is tuned at every step.

The probability of generating a configuration ${\cal C}_K = \{ {\vec r}_1, {\vec r}_2, \cdots , {\vec r}_K \}$ in the flatIGW algorithm is given by
\begin{equation}
P_N^{IGW}({\cal C}_K) = \prod _{L=1}^K p_L^{IGW}
\label{eq:pfIGW}
\end{equation}
Then, the variable ${\bar g}_K(m_K)$ is updated with the weight corresponding to ${\cal C}_K$:
\begin{equation}
{\bar g}_K(m_K) \longrightarrow {\bar g}_K(m_K) + [P_K^{IGW}({\cal C}_K)]^{-1}\ ; \quad
                                    m_K = \sum _{L=1}^K m({\vec r}_L)
\label{eq:wfIGW}
\end{equation}
We obtain an estimate, $g_K(m_K)$, of the DoS by averaging ${\bar g}_K(m_K)$ over all the attempts made to generate $K$-step walks. 

\section{Results and Discussion}

\begin{figure}
\centering
\includegraphics[width=2.5in,height=2.50in]{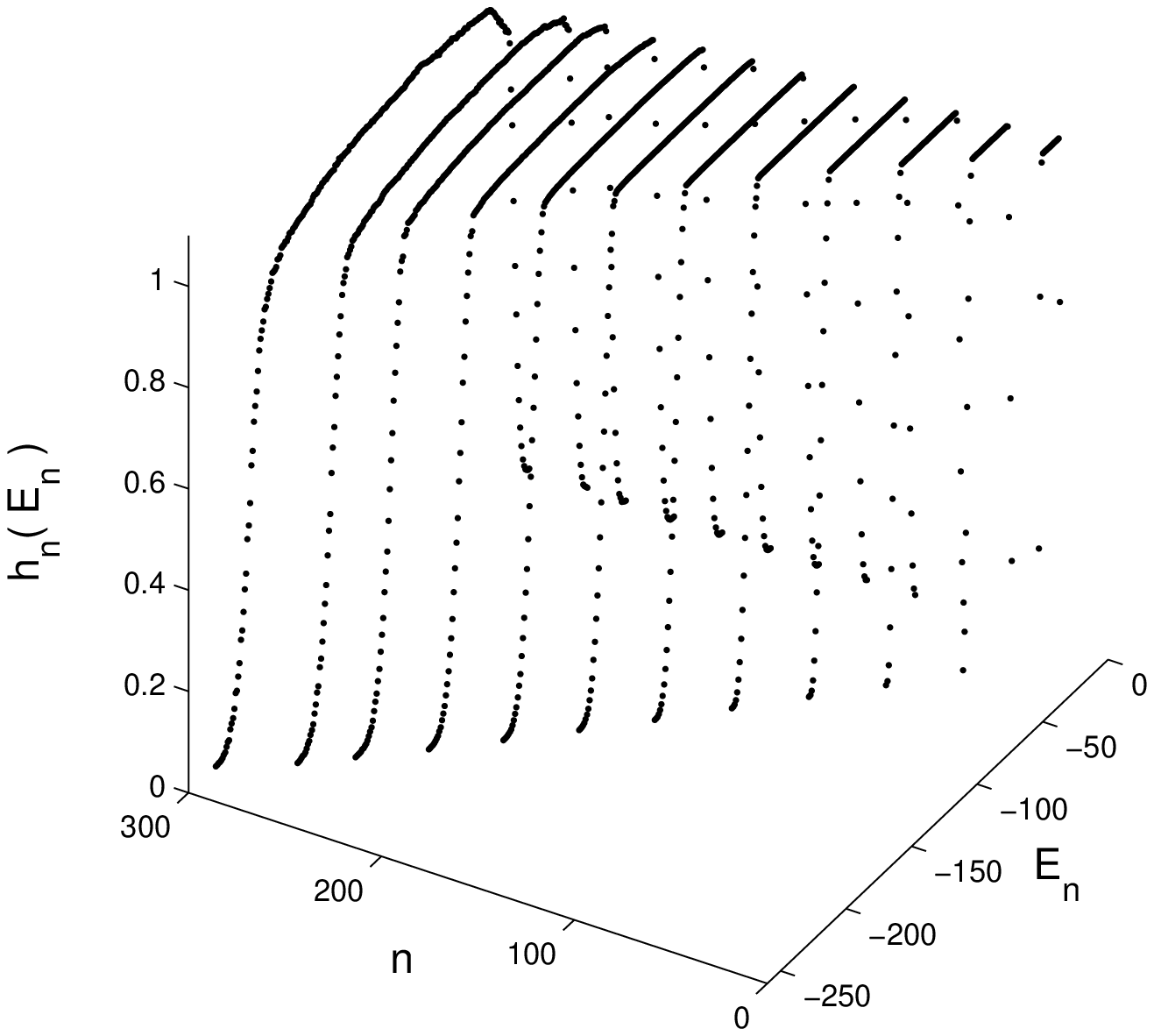}
\caption{Plot of reduced histogram $h_n(E_n)$ versus energy $E_n$ and length 
$n=25$ to $300$ (insteps of $25$: right to left) for ISAW.  Energy $-E_n$ is 
just the number of contacts made by a walk of length $n$.
These have been  obtained from flat
histogram  IGW on a square lattice with 
$\beta_G^{max}=1.0$ and $M=10^8$ attempts.}
\label{flathistsqr}
\end{figure}

\begin{figure}
\centering
\includegraphics[width=2.5in,height=2.5in]{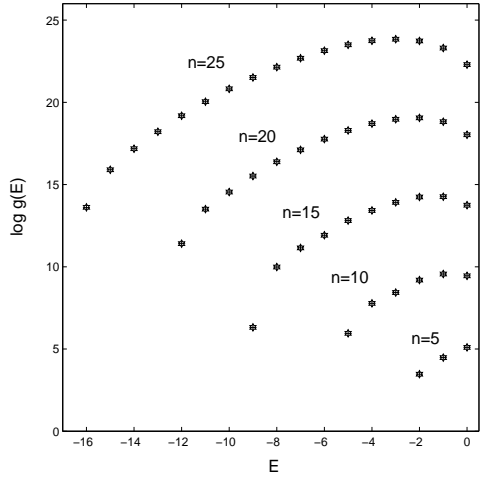}
\caption{Semilog plot of estimated  density of states $g^{est}_n(E_n)$ versus
energy $E_n$ and length $n=5$ to $25$ (insteps of $5$)  
for ISAW on a square lattices obtained from flatIGW with $\beta_G^{max}=1.0$; number of attempts $M=10^8$.
Energy $-E_n$ is just the number of contacts made by a walk of length $n$.
The Monte Carlo results ($\star$) matches  very well 
with exact (+) density of states. ( Monte Carlo errors are very small; maximum of error
does not exeed one percent and are within the symbol size on the scale).}
\label{flatdoscmp}
\end{figure}

\begin{figure}
\centering
\includegraphics[width=2.5 in,height=2.5in]{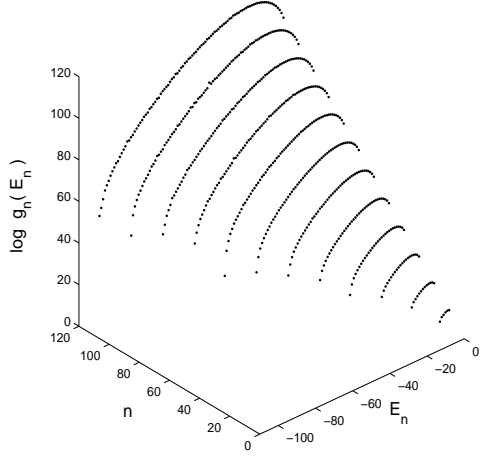}
\caption{Semilog plot of estimated  density of states $g^{est}_n(E_n)$ versus
energy $E_n$ and length $n=10$ to $120$ (insteps of $10$: right to left)
for ISAW on a square lattice obtained from flatIGW with $\beta_G^{max}=1.0$; number of attempts 
$M=10^8$. Energy $-E_n$ is just the number of contacts made by a walk of length $n$.}
\label{flatdossqr}
\end{figure}

\begin{figure}
\centering
\includegraphics[width=2.5in,height=3.5in]{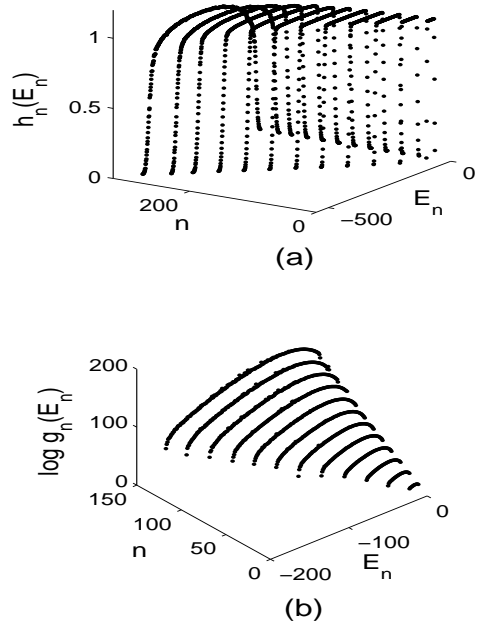}
\caption{Flat histogram IGW on a triangular lattice. (a) Plot of reduced histogram 
(b) Semilog plot of $g_n^{est}(E_n)$  vs $E_n$ for various $n$;
$\beta_G^{max}=0.8$; number of attempts $M=10^8$. Energy $-E_n$ is just the number of contacts made by a walk of length $n$.}
\label{flattri} 
\end{figure}

In Fig.(\ref{flathistsqr}), we show the reduced histogram, $h_K(m) = H_K(m)/$max$\{ H_K(m) \}$, for walks of  lengths in the range $K = 25$ to $300$ generated on a square lattice. The value of $\beta _G^{max}$ has been set equal to unity so that the adjustable parameter $\beta _G^{(L)}\in [-1, 1]$. The flat histograms seen are evidence that the local criterion used for adjusting $\beta _G^{(L)}$, Eq.(\ref{eq:delta} and Eq.(\ref{eq:dbeta4}), does bring about global flatness to the histogram data. The observation, Fig.(\ref{flatdoscmp}), that estimates of the DoS for short walks agree well with those obtained by exact enumeration gives confidence that the inverse algorithm, flatIGW, is as good as flatPERM. DoS estimated for longer walks, $K$ in the range $10$ to $120$ are shown in Fig.(\ref{flatdossqr}). Reduced histogram and the DoS estimates obtained, with $\beta _G^{max} = 0.8$, for walks on a triangular lattice are shown in Fig.(\ref{flattri}). Initially, a few trial runs may be necessary to fix suitable values of $\beta _G^{max}$ for different lattices.

Estimates of the DoS can be used for computing the specific heat of ISAW at any given bath temperature and, in general, for computing the phase diagram. ISAW is known to undergo a transition from random coil phase to $\theta$-phase and finally to a compact phase at temperatures that depend on the lattice used. So, locating a transition temperature, say the $\theta$-temperature, will be a test for the accuracy of DoS estimates. 

An interesting technique for locating a phase transition is to compute Fisher zeros of the canonical partition function, and check the (real) temperature to which they would converge. Since the walks generated are all of finite lengths, it is necessary to extrapolate the estimated temperatures to the asymptotic limit. 

\begin{figure}   
\centering
\includegraphics[width=2.16in]{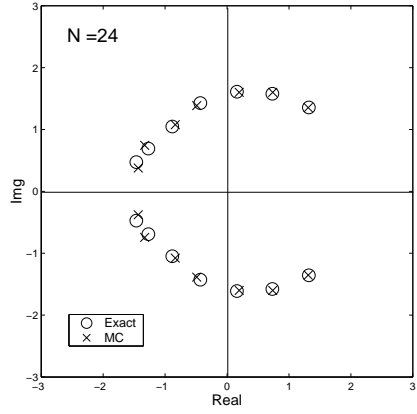}
\caption{Distribution of the roots of partition function of ISAW for $N=24$
on a square lattice. The results  obtained from exact enumeration (o) and  
flatIGW with $\beta_G^{max}=1.0$ (x). We have not marked 
two of the  roots since they lie far away from origin.}
\label{pzN24}
\end{figure}

\begin{figure}   
\centering
\includegraphics[width=3.16in,height=3.16in]{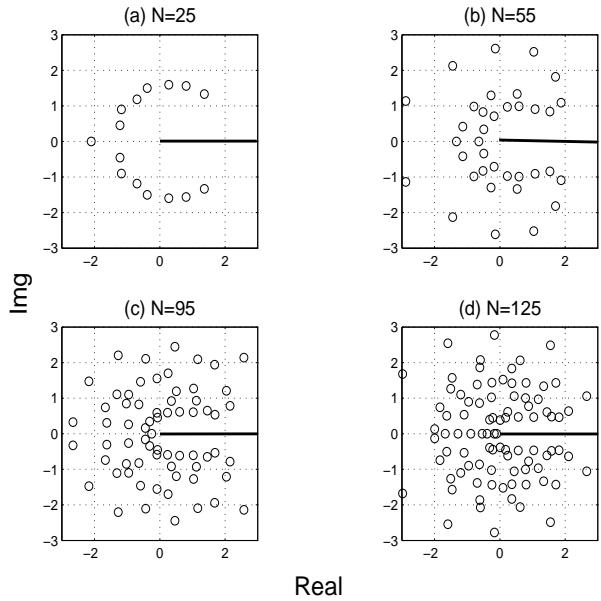}
\caption{Distribution of the roots of partition function of ISAW
on a square lattice for (a) $N=25$ (b) $N=55$ (c) $N=95$ (d) $N=125$.
The results  obtained from  flatIGW with $\beta_G^{max}=1.0$.
Note that some roots, lying too far away from the origin,
have been omitted.}
\label{pzflat}
\end{figure}

\begin{figure}   
\centering
\includegraphics[width=2.16in]{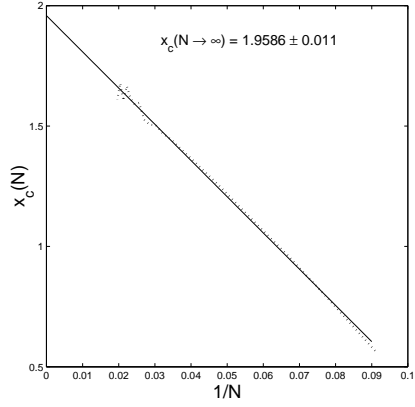}
\caption{ Results  obtained from  flatIGW  on a square lattice with 
$\beta_G^{max}=1.0$.
An extrapolation of the real part of leading zero, $x_c (N)$,
for $N=11$ to $51$ having the smallest magnitude imaginary component near 
positive real axis. The intercepts of the linear fit gives   
$x_c (N \to \infty ) = 1.9586 \pm 0.011$.}
\label{pzintercept} 
\end{figure}

Zeros of the polynomial, $Z_N$, in the variable $x(N) (\equiv e^{\beta})$ with the flatIGW estimates of the DoS as the coefficients, obtained for square lattice walks of length $N=24$, are shown in Fig.(\ref{pzN24}) along with those obtained by exact enumeration of $24$-step SAWs on a square lattice. The agreement is quite good. As the walk length increases, it is expected that there will be a crowding of zeros in the neighborhood of the positive real axis; Fig.(\ref{pzflat}) demonstrates this crowding of zeros for walks of lengths $N = 25, 55, 95$ and $125$. Numerical estimates of $x_c(N)$, the value at which a real zero is expected, obtained for various values of $N$ are plotted as function of $1/N$ in Fig.(\ref{pzintercept}). A simple extrapolation of the data gives the asymptotic value $x_c(N\to \infty) = 1.96 \pm 0.01$. This implies that $\beta _{\theta} \approx 0.67$ which is in close agreement with the known value $\sim 0.667$ \cite{barkema}. 

These results indicate that flatIGW can also be used for studying the statistical and thermodynamical properties of ISAW on a regular or disordered lattice. Recognizing the fact that the set of numbers, $\{ H_K(m)\}$, evolve as a rough surface, $\beta _G^{max}$ may be thought of as bounding lines within which the fluctuations of the $H$-surface are confined. So, it is intuitively clear that the $H$-surface will asymptotically reach a stationary 'flat' state; the time required for reaching this state and the asymptotic 'flatness' will depend on $\beta _G^{max}$.

flatIGW is an inverse algorithm in the following sense. The growth-rule of the flatIGW consists of two parts - (i) {\it local} IGW rule for sampling the available nearest neighbor sites and, (ii) {\it global} rule requiring that the  energy-histogram evolves in such a way as to maintain local flatness. The latter is used for adjusting the parameter, $\beta _G$, on which the former depends. So, these two parts of the growth-rule are inter-dependent. An indicator that this algorithm works is the (statistical) convergence of DoS towards their stationay values, as the weights assigned to the walks are accumulated independently during the simulation runs. In the direct methods, on the contrary, an indicator that the DoS converges to their stationary values is flattening of the energy-histogram during simulation; moreover, the {\it local} and the {\it global} components of the growth rule are independent.

flatIGW is an inverse algorithm similar in spirit to the Reverse Monte Carlo (RMC) algorithm of McGreevy and Pusztai \cite{McGreevy}, in which the model-parameters are adjusted so as to be consistent with the experimental data rather than by the standard Metropolis criterion.

However, the local flatness criterion, used in the step-growth rule, is not such a rigid constraint that a perfectly flat energy-histogram may be expected asymptotically. For instance, when $\mid \delta H_{K+1}(m_K+1)\mid = 0$ for all the available sites, $\beta _G^{K+1} = 0$ and therefore one of the sites will be chosen at random with equal probability; this will spoil the local flatness.

It must be mentioned here that the rule, Eq.(\ref{eq:delta}), used for adjusting $\beta _G$ in the algorithm is arbitrary and introduces one more parameter, $\beta _G^{max}$, besides using a randomizing factor $\xi$. This will affect the computational efficiency of the algorithm when compared to the other ones such as flatPERM \cite{flatperm} and Multi-canonical chain growth algorithm \cite{janke}. Nevertheless, the main objective of this work is to demonstrate that an inverse, histogram-driven algorithm ({\it a la} RMC) can be constructed for computing the DoS of linear polymers. A detailed comparative study of this algorithm {\it vis-a-vis} others will be helpful.

\section{Summary}

We have demonstrated that an inverse algorithm, flatIGW, that is similar in spirit to the Reverse Monte Carlo algorithm \cite{McGreevy} can be implemented for estimating the DoS of linear lattice polymers. This algorithm generates a SAW step-by-step and irreversibly by using the IGW growth rule, Eq.(\ref{eq:pIGW}), for a biased sampling of the available nearest neighbor sites; whether a contact-making available site is preferred or not is decided by a parameter, $\beta _G$, whose value is decided by the requirement of local flatness for the evolving energy-histogram.
The DoS converge to their stationary values, as the statistical weights associated with the walk configurations accumulate independently during the simulation runs. 
 
We have used this inverse algorithm to estimate the DoS for SAW on a square, as well as triangular, lattice. We have located the $\theta$ temperature for square lattice SAWs by analyzing the Fisher zeros corresponding to their DoS estimates. Agreement with the known value lends numerical support to this idea of an inverse algorithm. 

\bigskip

\noindent $^{\#}$Author for correspondence.

\newpage


\end{document}